\documentclass[12pt]{iopart}
\usepackage{iopams}
\usepackage{graphicx}
\usepackage{amssymb}
\usepackage{psfrag}
\begin{document}

\title{Black holes thermodynamics to all orders in the Planck length in
extra dimensions}
\author{Khireddine Nouicer\footnote%
{permanent address: Laboratory of Theoretical Physics,Faculty of
Sciences, University of Jijel, BP 98 Ouled Aissa, 18000 Jijel,
Algeria}}
\address{Frankfurt Institute of Advanced Studies,\\
John Wolfgang Goethe Universit\"at,\\
Max-von-Laue Str. 1, 60438 Frankfurt am Main, Germany}
\eads{nouicer@fias.uni-frankfurt.de}

\begin{abstract}
We investigate the effects to all orders in the Planck length, from
a generalized uncertainty principle (GUP), on the thermodynamic parameters
of radiating Schwarzschild black holes in a scenario with large
extra dimensions. We show that black holes in this framework are
hotter, have less degrees of freedom and decay faster compared to
black holes in the Hawking picture and in the framework with GUP to
leading order in the Planck length. Particularly, we show that the
final stage of the evaporation process is a black hole remnant with
zero entropy, zero heat capacity and non zero finite temperature. We
finally compare our results with the ones obtained in the standard
Hawking picture and with the generalized uncertainty principle to
leading order in the Planck length.
\end{abstract}


\maketitle
\section{Introduction}

The avenue of large extra dimension models (LED) offers new exciting ways
to solve the hierarchy problem and to study low scale quantum gravity
effects. The model of Alkanin-Hamed, Dimopoulos and Dvali (ADD) \cite%
{hamad1,hamad2,hamad3} used $n$ new large spacelike dimensions
without curvature, and gravity is the only force which propagates in
the full volume of the space-time (the bulk). Hence the
gravitational force in the four-dimensional world (the brane)
appears weak compared to the other forces which do not propagate in
the extra dimensions. An alternative model proposed by Randall and
Sundrum (RS) used warped extra dimension with a non factorisable
geometry \cite{Randall1,Randall2}. In RS models gravity is diluted
by the strong curvature of the extra dimension. Within these models
the Planck scale is lowered to values soon accessible of order of a
TeV. Among the predicted effects, the experimental production of
black holes (BHs) at particle colliders such as the Large Hadronic
Collider (LHC) \cite{lhc} and the muon collider \cite{mc}, is one of
the most exciting possibility which
has received a great amount of interest \cite{bh1}-\cite{bh33}%
. The newly formed BH is expected to decay instantaneously on collider
detector time scales (typically of order $10^{-26} s$ for LHC). At this
scale the evaporation of BH is expected to end leaving up a possible
Planck size black hole remnant (BHR).

Recently, a great interest has been devoted to the study of effects
of generalized uncertainty principles (GUPs) and modified dispersion
relations (MDRs) on various quantum gravity problems
\cite{ma}-\cite{chris}. The GUPs and MDRs originates from several
studies in string theory approach to quantum gravity
\cite{ve}-\cite{koni}, loop quantum gravity \cite{ga},
noncommutative space-time algebra \cite{ma1,ke} and black holes
gedanken experiments \cite{ma2,sc}. Actually, GUPs and MDRs are
considered as common features of any promising candidate to a
quantum theory of gravity.

In four dimensions, the consequences of GUPs and/or MDRs on BHs
thermodynamics have been considered intensively in the recent
literature on the subject \cite{bo}-\cite{noa}, notably it has been
shown that GUP prevents black holes from complete evaporation
exactly like the standard
Heisenberg principle prevents the hydrogen atom from total collapse \cite%
{adler}. Then, at the final stage of the Hawking evaporation process of a black
hole, a inert black hole remnant (BHR) continue to exist with zero entropy,
zero heat capacity and a finite non zero temperature. The inert character of
the BHR, besides gravitational interactions, renders this object a serious
candidate to explain the origin of dark matter \cite{Pisin,P-adler}. A
particular attention has been also devoted to the computation of the entropy
and to the sub-leading logarithmic correction \cite{c12}-\cite{c24}%
. The phenomenological properties of Black holes in the framework of the ADD
model with GUP have been recently also studied \cite{bh25,noza}.

Until now all of the work has been done with GUP in the leading
order in the fundamental length. However, a version of the GUP with
higher orders in the Planck length induces quantitative corrections
to the entropy and then influences the Hawking evaporation of the
black hole \cite{ko}. Then, the ultimate quantum nature of the
physics at the Planck scale would be best described in the framework
of a GUP containing the gravitational effects to all orders in the
Planck length . In this framework, the corrections to BH
thermodynamic parameters may have important consequences on BHs
production at particle colliders.

In this paper we discuss the effects, to all orders in the Planck
length, that a GUP may have on thermodynamic parameters of the
Schwarzschild BH in the ADD model. The organization of this work is
as follows. In section 2, we introduce a deformed position and
momentum operators algebra leading to the GUP
and examine quantum properties of this algebra. In section 3, the GUP%
-corrected thermodynamic parameters are computed and the departures from the
standard semiclassical description shown. In section 4, we investigate the Hawking evaporation
process and calculate exactly the evaporation rate and the decay time.
We compare our results with the ones obtained in the context of the semiclassical description and with the
GUP to the leading order in the Planck length. Our conclusions are summarized in the last section.

\section{All orders corrections of GUP}

One of the most interesting consequences of all promising quantum
gravity candidates is the existence of a minimal observable length
on the order of the Planck length. Actually, part of the work in quantum gravity phenomenology has been tackled
with effective models based on MDRs and/or GUPs and containing the
minimal length as a natural UV cut-off. The relation between these
approaches has been recently clarified and established
\cite{Sabine}.

The idea of a minimal length can be modeled in terms of a quantized
space-time and goes back to the early days of quantum field theory
\cite{snyder} (see also $\cite{connes}-\cite{bondia}$ ). An
alternative approach is to
consider deformations to the standard Heisenberg algebra \cite{ke}%
, which lead to  generalized uncertainty principles showing the
existence of the minimal length. In this section we follow the
latter approach and exploit a result recently obtained in the
context of canonical noncommutative field theory in the coherent
states representation \cite{spallucci} and field theory on
non-anticommutative superspace \cite{Moffat,nouicer}. Indeed, it has
been shown  that the Feynman propagator displays an exponential UV
cut-off of the form $\exp \left( -\eta p^{2}\right) $, where the
parameter $\eta $ is related to the minimal
length. This framework has been further applied, in series of papers \cite%
{nico}, to the black hole evaporation process.

At the quantum mechanical level, the UV finiteness of the
Feynman propagator can be also captured by a non linear relation, $k=f(p)$,
between the wave vector and the momentum of the particle \cite{Sabine}. This
relation must be invertible and has to fulfil the following requirements:

\begin{enumerate}
\item For  energies much smaller than the cut-off the usual dispersion relation is
recovered.

\item The wave vector is bounded by the cut-off.
\end{enumerate}

In this picture, the usual commutator between the commuting position and momentum operators is generalized to
\begin{equation}
[X,P]=i\hbar\frac{\partial p}{\partial k}\Leftrightarrow \Delta X\Delta P\geq\frac{\hbar}{2}\left|\left\langle\frac{\partial p}{\partial k}\right\rangle\right|,
\end{equation}
and the momentum measure $d^{n}p$ is deformed as $%
d^{n}p\prod_{i}\frac{\partial k_{i}}{\partial p_{j}}$. In the
following, we will restrict ourselves to the isotropic case in one
space-like
dimension. Following \cite{spallucci,nouicer} and setting $\eta =\frac{%
\alpha L_{Pl}^{2}}{\hbar ^{2}}$ we have
\begin{equation}
\frac{\partial p}{\partial k}=\hbar  { \exp}\left( \frac{\alpha L_{Pl}^{2}%
}{\hbar ^{2}}p^{2}\right) ,  \label{measure}
\end{equation}%
where $\alpha $ is a dimensionless constant of order one.

From Eq.$\left( \ref{measure}\right) $ we obtain the dispersion relation
\begin{equation}
k\left( p\right) =\frac{\sqrt{\pi }}{2\sqrt{\alpha} L_{Pl}} {  erf}\left( \frac{%
\sqrt{\alpha }L_{Pl}}{\hbar }p\right) \label{mdr},
\end{equation}%
from which we have the following minimum Compton wavelength
\begin{equation}
\lambda _{0}=4\sqrt{\pi \alpha }L_{Pl}.  \label{compton}
\end{equation}
We note that a dispersion relation similar to the one given by
Eq.(3) has been used recently to investigate the effect of the
minimal length on the running gauge couplings \cite{sab}. In the
context of trans-Plankian physics, modified dispersion relations
have been also used to study the spectrum of the cosmological
fluctuations. A particular class of MDRs frequently used in the
literature \cite{slot, jora} is the well known Unruh dispersion
relations given by $k(p) = tanh^{1/\gamma}(p^{\gamma})$, with
$\gamma$ being some positive integer \cite{unruh}.

Let us show that the above results can be obtained from the
following momentum space representation of the position and momentum
operators \
\begin{equation}
X=i\hbar \exp \left( \frac{\alpha L_{Pl}^{2}}{\hbar ^{2}}P^{2}\right) {%
\partial _{p}}\qquad P=p.  \label{xp}
\end{equation}
The corrections to the standard Heisenberg algebra become effective in the
so-called quantum regime where the momentum and length scales are of the
order of the Planck mass $M_{Pl}$ and  the Planck length $L_{Pl}$
respectively.

The hermiticity condition of the position operator implies  modified completeness relation and modified scalar product given by
\begin{equation}
\int dpe^{-\frac{\alpha L_{Pl}^{2}}{\hbar ^{2}}p^{2}}|p\rangle \langle p|=1
\label{ferm}
\end{equation}%
\begin{equation}
\left\langle p\right\vert \left. p^{\prime }\right\rangle =e^{\frac{\alpha
L_{ {Pl}}^{2}}{\hbar ^{2}}p^{2}}\delta \left( p-p^{\prime }\right) .
\end{equation}%
From Eq.$\left( \ref{ferm}\right) $, we observe that we have reproduced the
Gaussian damping factor in the Feynman propagator \cite{spallucci,nouicer}.

The algebra defined by Eq. $\left( \ref{xp}\right) $ leads to the following
generalized commutator and  generalized uncertainty principle (GUP)
\begin{equation}
\left[ X,P\right] =i\hbar \exp \left( \frac{\alpha L_{Pl}^{2}}{\hbar ^{2}}%
P^{2}\right) ,\quad \left( \delta X\right) \left( \delta P\right) \geq \frac{%
\hbar }{2}\left\langle \exp \left( \frac{\alpha L_{Pl}^{2}}{\hbar ^{2}}%
P^{2}\right) \right\rangle .  \label{GUP}
\end{equation}%

In order to investigate the quantum mechanical implications of this
deformed algebra, we solve the relation $\left(\ref{GUP}\right)$ for
$\left( \delta P\right) $ with the equality. Using the property
$\left\langle P^{2n}\right\rangle \geq \left\langle
P^{2}\right\rangle $ and $\left( \delta P\right) ^{2}=\left\langle
P^{2}\right\rangle -\left\langle P\right\rangle ^{2}$, the
generalized uncertainty relation is written as
\begin{equation}
\left( \delta X\right) \left( \delta P\right) =\frac{\hbar }{2}\exp \left(
\frac{\alpha L_{P {l}}^{2}}{\hbar ^{2}}\left( \left( \delta P\right)
^{2}+\left\langle  P\right\rangle ^{2}\right) \right) .
\end{equation}%
Taking the square of this expression we obtain
\begin{equation}
W\left( u\right) e^{W\left( u\right) }=u,  \label{lam},
\end{equation}%
where we have set $W(u)=-2\frac{\alpha L_{Pl}^{2}}{\hbar ^{2}}\left( \delta
P\right) ^{2}$ and $u=-\frac{\alpha L_{Pl}^{2}}{2\left( \delta X\right) ^{2}}%
e^{-2\frac{\alpha L_{P {l}}^{2}}{\hbar ^{2}}\left\langle P\right\rangle
^{2}}.$

The equation given by Eq.$\left(\ref{lam}\right)$ is exactly the definition
of the Lambert function \cite{Lambert}, which is a
multi-valued function. Its different branches, $W_k(u)$, are labeled by the integer $%
k=0,\pm 1,\pm 2,\cdots $. When $u$ is a real number Eq.$\left(\ref{lam}%
\right)$ have two real solutions for $0\geq u\geq -\frac{1}{e}$, denoted by $%
W_{0}(u)$ and $W_{-1}(u)$, or it can have only one real solution for $u\geq
0 $, namely $W_{0}(u)$ . For -$\infty <u<-\frac{1}{e}$, Eq.(\ref{lam}) have
no real solutions.

Finally, the momentum uncertainty is given by
\begin{equation}
\left( \delta P\right) =\frac{\hbar}{\sqrt{2\alpha}L_{Pl}}\left(-W\left(-\frac{\alpha L_{Pl}^{2}}{2\left( \delta X\right) ^{2}}%
e^{-2\frac{\alpha L_{P {l}}^{2}}{\hbar ^{2}}\left\langle P\right\rangle
^{2}}\right)\right)^{1/2}.  \label{argu}
\end{equation}%
From the argument of the Lambert function we have the following condition
\begin{equation}
\frac{\alpha L_{Pl}^{2}e^{\frac{2\alpha L_{Pl}^{2}}{\hbar ^{2}}%
\left\langle P\right\rangle ^{2}}}{2\left( \delta X\right) ^{2}}\leqslant
\frac{1}{e},
\end{equation}%
which leads to a minimal uncertainty in position given by
\begin{equation}
\left( \delta X\right) _{\min }=\sqrt{\frac{e\alpha}{2}} L_{Pl}e^{\frac{%
\alpha L_{Pl}^{2}}{\hbar ^{2}}\left\langle P\right\rangle ^{2}}.
\end{equation}%
The absolutely smallest uncertainty in position or minimal length \ is
obtained for physical states for which we have $\left\langle P\right\rangle
=0$ and $\left( \delta P\right) =\hbar /\left( \sqrt{2\alpha }L_{P {l}%
}\right) ,$ and is given by
\begin{equation}
\left( \delta X\right) _{0}=\sqrt{\frac{\alpha e}{2}}L_{Pl}.  \label{min}
\end{equation}%
In terms of the minimal length the momentum uncertainty becomes
\begin{equation}
\left( \delta P\right) =\frac{\hbar\sqrt{e}}{2(\delta X)_0}\left( -%
W\left( -\frac{1}{e}\left( \frac{(\delta X)_{0}}{(\delta X)}%
\right) ^{2}\right) \right)^{1/2} .  \label{argup}
\end{equation}%
Here we observe that $\frac{1}{e}\frac{(\delta X)_{0}}{(\delta
X)}<1$ is a small parameter by virtue of the GUP, and perturbative
expansions to all orders in the Planck length can be safely
performed.

Indeed, a series expansion of Eq.(\ref{argup}) gives the corrections to the
standard Heisenberg principle
\begin{equation}
\delta P\simeq \frac{\hbar }{2\left( \delta X\right) }\bigg( 1+\frac{1}{2e}%
\left( \frac{(\delta X)_{0}}{(\delta X)}\right) ^{2}+\frac{5}{8e^{2}}\left(
\frac{(\delta X)_{0}}{(\delta X)}\right) ^{4}
+\frac{49}{48e^{3}}\left( \frac{%
(\delta X)_{0}}{(\delta X)}\right) ^{6}+\ldots \bigg) .
\end{equation}%
This expression of $\left( \delta P\right) $ containing only odd powers of $%
\left( \delta X\right) $ is consistent with a recent analysis in which
string theory and loop quantum gravity, considered as the most serious
candidates for a theory of quantum gravity, put severe constraints on the
possible forms of GUPs and MDRs \cite{nozari}.

Let us now recall the form of the GUP to leading order in the Planck
length. This GUP is given by
\begin{equation}
\left( \delta X\right) \left( \delta P\right) \geq \frac{\hbar }{2}\left( 1+%
\frac{\alpha L_{Pl}^{2}}{\hbar ^{2}}\left( \delta P\right) ^{2}\right) .
\label{gup}
\end{equation}%
A simple calculation leads to the following minimal length
\begin{equation}
\label{l0}
\left( \delta X\right) _{0}=\sqrt{\alpha }L_{Pl},
\end{equation}%
which is of order of the Planck length.  However, the form of the
GUP to leading order in the Planck length leads to a modified
dispersion relation which does not fulfill
the second requirement listed above \cite%
{Sabine}. In our case, It is easy to show that the wave vector given
by $(\ref{mdr})$ is bounded by the cut-off $1/L_{Pl}$. This
observation may significantly influence the thermodynamics
parameters and the evaporation process of small BHs.

In the following sections we use the form of the GUP given by Eq.$%
\left( \ref{GUP}\right) $ and investigate the thermodynamics of
the Schwarzschild BH. We use units $\hbar =c=k_{ {B}}=1$.

\section{Black hole thermodynamics with GUP}

Black holes in higher dimensional space-times have been studied by
Myers and Perry \cite{perry}. They considered the form of the
gravitational background around an uncharged $d$-dimensional BH. In
the non-rotating case, corresponding to a $d$-dimensional
spherically symmetric Schwarzschild BH, the line element is given by
\begin{eqnarray}
ds^{2}&=&-\left( 1-\frac{16\pi G_{d}M}{\left( d-2\right) \Omega _{d-2}r^{d-3}}%
\right) dt^{2}-\left( 1-\frac{16\pi G_{d}M}{\left( d-2\right) \Omega
_{d-2}r^{d-3}}\right) ^{-1}dr^{2}\nonumber \\
&-&r^{2}d\Omega _{d-2}^{2}.
\end{eqnarray}%
where $\Omega _{d-2}$ is the metric of the unit sphere $S^{d-2}$ and
$G_{d}=G_4 L^{d-4}$ is the $d$-dimensional Newton's constant and $L$
the size of the extra dimensions. The horizon radius $r_{\hbox{h}}$
is defined by the vanishing of the component $g_{00}$ and is given
by

\begin{equation}
r_{\hbox{h}}=\left( \frac{16\pi G_{d}M}{\left( d-2\right) \Omega _{d-2}}%
\right) ^{\frac{1}{d-3}}=\omega _{d}L_{Pl}m^{\frac{1}{d-3}}  \label{r0}
\end{equation}%
with
\begin{equation}
\omega _{d}=\left( \frac{16\pi }{\left( d-2\right) \Omega _{d-2}}\right) ^{%
\frac{1}{d-3}},\quad m=\frac{M}{M_{Pl}},
\end{equation}%
and $M_{Pl}=\left( {G_{d}}\right) ^{-\frac{1}{d-3}}$ is the
fundamental $d$-dimensional Planck mass. From Eq.(\ref{r0}) we
observe that the horizon radius increases with the space-time
dimension, reflecting the strong gravity effects at small distances.

In the standard case, the Hawking temperature and entropy of a BH of
large mass $M$ are given by \cite{bh25}
\begin{equation}
T_{H}=\frac{d-3}{4\pi \omega _{d}m^{1/(d-3)}}M_{Pl},\quad
S=\frac{d-3}{4L_{Pl}^2}A,
\end{equation}%
where $A=\Omega_{d-2}r_h^{d-2}$ is the BH horizon area.

Let us then examine the effects to all orders in the Planck length brought by the GUP defined by $\left( \ref{lam}%
\right) $, on the Hawking temperature and entropy
. Following the heuristic argument of \cite{adler}, based on
the uncertainty principle, we have
\begin{equation}
T_{H}=\frac{\left( d-3\right) \delta P}{2\pi }.
\end{equation}%
In our framework, we use  (\ref{lam}) and consider BH
near geometry for which $\delta X\simeq r_{h}=\omega _{d}L_{Pl}%
m^{\frac{1}{d-3}}.$ In this case, the existence of the minimal length leads
to the following non zero BH minimum mass
\begin{equation}
m_{0}=\left( \frac{\left( \delta X\right) _{0}}{\omega _{d}L_{Pl}}\right)
^{d-3}=\left( \sqrt{\frac{e}{2}}\frac{\alpha }{\omega _{d}}\right) ^{d-3}.
\label{min-mass}
\end{equation}%
It is interesting to note that the BH minimum mass presents a
maximum for $d=12$, and tends asymptotically to zero for $d\geq 30$.
In terms of the BH mass, the GUP-corrected Hawking temperature is
\begin{equation}
T_{H}=\frac{\left( d-3\right) T_{Pl}}{4\pi \omega _{d}m^{\frac{1}{d-3}%
}}\exp \left( -\frac{1}{2}W\left( -\frac{1}{e}\left( \frac{m_{0}}{m}\right)
^{2/\left( d-3\right) }\right) \right) .  \label{H-temp}
\end{equation}%
From this expression we observe, that the BH temperature is only defined for $m\geq m_0$.
For a BH with a mass equal to $m_{0}$, the Hawking temperature reaches a maximum given by%
\begin{equation}
T_{H}^{\max }=\frac{(d-3)}{2\sqrt{2}\pi \alpha }T_{Pl}.  \label{Tmax}
\end{equation}%
The corrections to the standard Hawking temperature are obtained by
expanding $\left( \ref{H-temp}\right) $ in terms of $\frac{1}{e}\left( m_{0}/%
{m}\right) ^{2/\left( d-3\right) }$
\begin{equation}
T_{H}\approx \frac{\left( d-3\right) T_{Pl}}{4\pi \omega _{d}m^{\frac{%
1}{d-3}}}\left[ 1+\frac{1}{2e}\left( \frac{m_{0}}{m}\right) ^{2/(d-3)}+\frac{%
5}{8e^{2}}\left( \frac{m_{0}}{m}\right) ^{4/(d-3)}+\cdots \right] .
\end{equation}%
In the limit $m_{0}=0$, the standard expression is recovered.
However, as mentioned above $m_{0}\rightarrow0$ for larger $d$, and in this case the standard expression of the BH
temperature is also reproduced. This means that BHs  with GUP in
higher dimensional space-time evaporate completely exactly like in the semiclassical picture.

In figure 1, we show the variation of the corrected Hawking
temperature ( \ref{H-temp})  with the BH mass. We observe that BHs
in a scenario with extra dimensions are hotter and consequently
tends to evaporate faster.
\begin{center}
\includegraphics[height=8cm,
width=10cm]{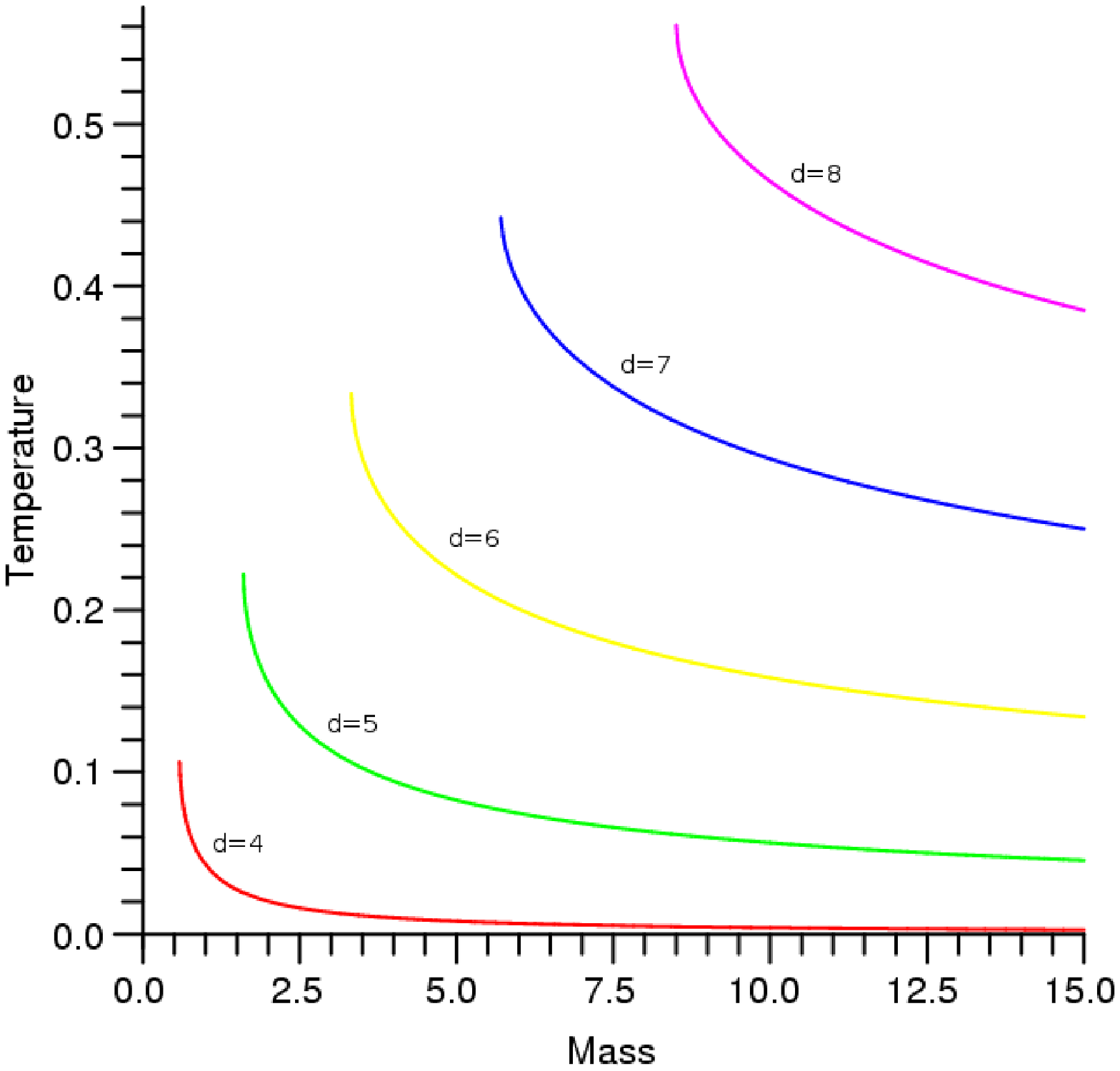}
\end{center}
\begin{center}
{Figure 1: Corrected Hawking temperature versus the black hole mass
(in units with $G=L=1$).}
\end{center}

We turn now to the calculation of the micro-canonical entropy of a
large BH. Following heuristic considerations due to Bekenstein, the
minimum increase of the area of a BH absorbing a classical particle
of
energy $E$ and size $R$ is given by $\left( \Delta A\right) _{min%
}\simeq \frac{4\left( \ln 2\right) L_{Pl}^{d-2}ER}{\left( d-3\right) }$. At
the quantum mechanical level, the size and the energy of the particle are
constrained to verify $R\sim 2\delta X$ and $E\sim \delta P$. Then we have $%
\left( \Delta A\right) _{min}\simeq \frac{8\left( \ln 2\right) L_{%
Pl}^{d-2}\delta X\delta P}{\left( d-3\right) }.$
Extending this approach to the case with GUP  we obtain
\begin{equation}
\left( \Delta A\right) _{min}\approx \frac{4\left( \ln 2\right)
L_{Pl}^{d-2}}{\left( d-3\right) }\exp \left( -\frac{1}{2}W\left( -\frac{1}{e}%
\left( \frac{A_{0}}{A}\right) ^{2/(d-2)}\right) \right) ,
\end{equation}
where $A=\Omega _{d-2}\left( r_{h}\right) ^{d-2}$ and $A_{0}=\Omega
_{d-2}\left( \delta X\right) _{0}^{d-2}$ are respectively the BH horizon
area and minimum horizon area. Considering near horizon
geometry,  for which we have $\delta X$ as the horizon radius, and with the aid of the Bekenstein
calibration factor for the minimum increase of entropy $\left( \Delta
S\right) _{min}=\ln 2$, we have
\begin{equation}
\frac{dS}{dA}\simeq \frac{5\left( \Delta S\right) _{min}}{\left(
\Delta A\right) _{min}}=\frac{\left( d-3\right) }{4L_{Pl}^{d-2}}\exp
\left( \frac{1}{2}W\left( -\frac{1}{e}\left( \frac{A_{0}}{A}\right)
^{2/(d-2)}\right) \right) .
\end{equation}
Then , up to an irrelevant constant, we write the entropy as
\begin{equation}
S_{d}=\frac{\left( d-3\right) }{4L_{Pl}^{d-2}}\int_{A_{0}}^{A}\exp
\left( \frac{1}{2}W\left( -\frac{1}{e}\left( \frac{A_{0}}{A}\right)
^{2/(d-2)}\right) \right) dA.
\end{equation}%
The lower limit of integration is a consequence of the GUP.
Using the variable $y=\frac{1}{e}\left( \frac{A_{0}}{A}\right)
^{2/(d-2)}$ and the relation $e^{\frac{W(x)}{2}}=\sqrt{x/W(x)}$ we have
\begin{equation}
S_{d}=-\frac{\left( d-3\right) \left( d-2\right) }{8\left( \sqrt{e}%
L_{Pl}\right) ^{d-2}}A_{0}\int_{\frac{1}{e}}^{\frac{1}{e}\left( \frac{A_{0}}{%
A}\right) ^{2/(d-2)}}y^{-\frac{d}{2}}\left[ \frac{-y}{W\left( -y\right) }%
\right] ^{\frac{1}{2}}dy,
\end{equation}
Performing the integration we finally obtain the following corrected
BH entropy for some values of $d$.  Up to a constant, which is the
value of the entropy for $ y=1/e$, we obtain for $d=4:$
\begin{equation}
S_{4}=\frac{A_{0}}{8\left( \sqrt{e}L_{Pl}\right) ^{2}}\left[ \frac{2}{\sqrt{%
zW\left( z\right) }}-Ei\left( 1,\frac{1}{2}W\left( z\right) \right) %
\right] _{z=-\frac{1}{e}\left( A_{0}/A\right) },
\end{equation}%
$d=5:$
\begin{eqnarray}
S_{5}&=&-\frac{A_{0}}{2\left( \sqrt{e}L_{Pl}\right) ^{3}z\sqrt{-W\left(
z\right) }}\bigg[ 1+W\left( z\right) \nonumber \\
&+&\sqrt{\pi }z\sqrt{W\left( z\right) }%
erf\left( \sqrt{W\left( z\right) }\right) \bigg] _{{z=-\frac{1}{e}\left( A_{0}/A\right)^{2/3} }},
\end{eqnarray}%
$d=6:$
\begin{eqnarray}
S_{6}&=&\frac{3A_{0}}{8\left( \sqrt{e}L_{Pl}\right) ^{4}}\bigg[- \frac{2}{z%
\sqrt{zW\left( z\right) }}\nonumber \\
&-&\frac{3}{2}Ei\left( 1,\frac{3}{2}W\left(
z\right) \right) -\frac{1}{z}\sqrt{\frac{W\left( z\right) }{z}}\bigg]
_{{z=-\frac{1}{e}\left( A_{0}/A\right)^{1/2} }},
\end{eqnarray}%
$d=7:$
\begin{eqnarray}
S_{7}&=&\frac{A_{0}}{3\left( \sqrt{e}L_{Pl}\right) ^{5}z^{2}\sqrt{-W\left(
z\right) }}\bigg[ W\left( z\right) -4\left( W\left( z\right) \right) ^{2}\nonumber\\
&-&4
\sqrt{2\pi }z^{2}\sqrt{W\left( z\right) }erf\left( \sqrt{2}W\left(
z\right)\right)+3\bigg]_{{z=-\frac{1}{e}\left( A_{0}/A\right)^{2/5} }},
\end{eqnarray}%
$d=8:$
\begin{eqnarray}
S_{8}&=&\frac{5A_{0}}{64\left( \sqrt{e}L_{Pl}\right) ^{6}}\bigg[
10\left( \frac{W\left( z\right) }{z}\right) ^{\frac{3}{2}}-%
\frac{4}{z^{2}}\sqrt{\frac{W\left( z\right) }{z}}-\frac{16}{%
z^{2}\sqrt{zW\left( z\right) }}\nonumber\\
&-&25Ei\left( 1,\frac{5}{2}W\left(
z\right) \right) \bigg]_{z=-\frac{1}{e}\left( A_{0}/A\right)^{1/3} },
\end{eqnarray}%
where $erf\left( z\right)$ is the error function and $ Ei\left( n,z\right)
$ is the exponential integral. The corrections
to the standard expressions are obtained by applying a Taylor expansion
around the parameter $z$ which is a small one by virtue of the GUP.
For $d=4$ we obtain
\begin{eqnarray}
S_{4}&=&\frac{A}{4L_{Pl}^{2}}-\frac{\pi \alpha ^{2}}{4}\ln \frac{A}{A_{0}}+%
\frac{\pi \alpha ^{3}}{16e}\left( \frac{A_{0}}{A}\right)\nonumber \\
&+&\frac{25\pi \alpha
^{2}}{192e^{2}}\left( \frac{A_{0}}{A}\right) ^{2}+\frac{343\pi \alpha ^{2}}{%
2304e^{3}}\left( \frac{A_{0}}{A}\right) ^{3}+\cdots,
\label{expd4}
\end{eqnarray}%
and for $d=5,6,7,8$ we have respectively%
\begin{eqnarray}
S_{5} &=&\frac{A}{2L_{Pl}^3}-\frac{3}{4\sqrt{2}}\pi^2\alpha^3 e^{1/2}\left(\frac{A}{A_0}\right)^{1/3}
+\frac{9}{16\sqrt{2}}\frac{\pi^2\alpha^3}{e^{1/2}}\left(\frac{A_0}{A}\right)^{1/3}\nonumber \\
&+&\frac{25}{96\sqrt{2}}\frac{\pi^2\alpha^3}{e^{3/2}}\left(\frac{A_0}{A}\right)
+\frac{196}{1280\sqrt{2}}\frac{\pi^2\alpha^3}{e^{5/2}}\left(\frac{A_0}{A}\right)^{5/3}
+\cdots,  \label{d5}
\end{eqnarray}
\begin{eqnarray}
S_{6} &=&\frac{3A}{4L_{Pl}^4}-\frac{\pi^2 e\alpha^4}{2}\left(\frac{A}{A_0}\right)^{1/2}
-\frac{3}{16}\pi^2\alpha^4\ln\frac{A}{A_0}+\frac{25}{48}\frac{\pi^2\alpha^4}{e}\left(\frac{A_0}{A}\right)^{1/2} \nonumber \\
&&+\frac{343}{768}\frac{\pi^2\alpha^4}{e^2}\left(\frac{A_0}{A}\right)+\frac{727}{1280}\frac{\pi^2\alpha^4}{e^3}
\left(\frac{A_0}{A}\right)^{3/2}+\cdots, \label{d6}
\end{eqnarray}
\begin{eqnarray}
S_{7} &=&\frac{A}{L_{Pl}^{5}}-\frac{5}{24\sqrt{2}}\pi^3\alpha^5 e^{3/2}\left(\frac{A}{A_0}\right)^
{3/5}-\frac{15}{32\sqrt{2}}\pi^3\alpha^5 e^{1/2}\left(\frac{A}{A_0}\right)^
{1/5}\nonumber \\
&+&\frac{125}{192\sqrt{2}}\frac{\pi^3\alpha^5}{e^{1/2}}\left(\frac{A_0}{A}\right)^
{1/5}
 +\frac{1715}{4608\sqrt{2}}\frac{\pi^3\alpha^5}{e^{3/2}}\left(\frac{A_0}{A}\right)^
{3/5}+\cdots,
\end{eqnarray}
\begin{eqnarray}
S_{8} &= &\frac{5A}{4L_{Pl}^{6}}-\frac{1}{8}\pi^3\alpha^6 e^2\left(\frac{A}{A_0}\right)^{2/3}
-\frac{3}{16}\pi^3\alpha^6 e\left(\frac{A}{A_0}\right)^{1/3}\nonumber \\
&-&\frac{25}{288}\pi^3\alpha^6\ln\frac{A}{A_0}+\frac{343}{768}\frac{\pi^3\alpha^6}{ e}\left(\frac{A_0}{A}\right)^{1/3}
+\frac{2187}{5120}\frac{\pi^3\alpha^6 }{e^2}\left(\frac{A_0}{A}\right)^{2/3}+\cdots .
\end{eqnarray}
We note that we have reproduced in the case with even number of dimensions
the log-area correction term with a negative sign, since we are dealing with the micro-canonical entropy.
For $d=4$, the expansion coefficients are proportional to $\alpha ^{2\left( n+1\right) }$, exactly as in \cite{me1}.

In order to analyze the question of how a generalization of the Heisenberg uncertainty principle might influence the BH entropy and then the BH decay, we construct the following ratio between the entropy calculated in different scenarios
\begin{equation}
R_0=\frac{S_{ao}  }{S_{H}},\qquad R_1=\frac{S_{ao}}{S_{lo}},
\end{equation}
where $S_{ao}, S_{lo}, S_{H}$ are respectively the entropy with the GUP to all orders in the Planck length, the entropy with GUP to leading order in the Planck length (see the end of section 4), and the entropy in the Hawking picture. The results of this analysis are shown in Figs. 2 and 3 where we see that $R_0$ and $R_1$ increase with $m$ and decrease with $d$. For large $m$, $R_0$ tends slowly to unity in comparison to $R_1$.  This shows, that the effects of  the GUP  become relevant as the mass of the BH decreases.

Regarding our results, we conclude that the BH entropy is smaller
than the ones obtained in the Hawking picture and with GUP to
leading order in the Planck length. On the other hand the BH entropy
decreases with the number of dimensions confirming the predictions
of GUP to leading order in the Planck length. This indicates that
BHs in scenario with extra dimensions and GUP to all orders in the
Planck length have less degrees of freedom compared to their
counterparts in the Hawking picture and GUP to leading order in the
Planck length. Then, in our framework we expect a significant
suppression of the multiplicity of the emitted particles in the
evaporation phase.

\begin{center}
\includegraphics[height=8cm,
width=10cm]{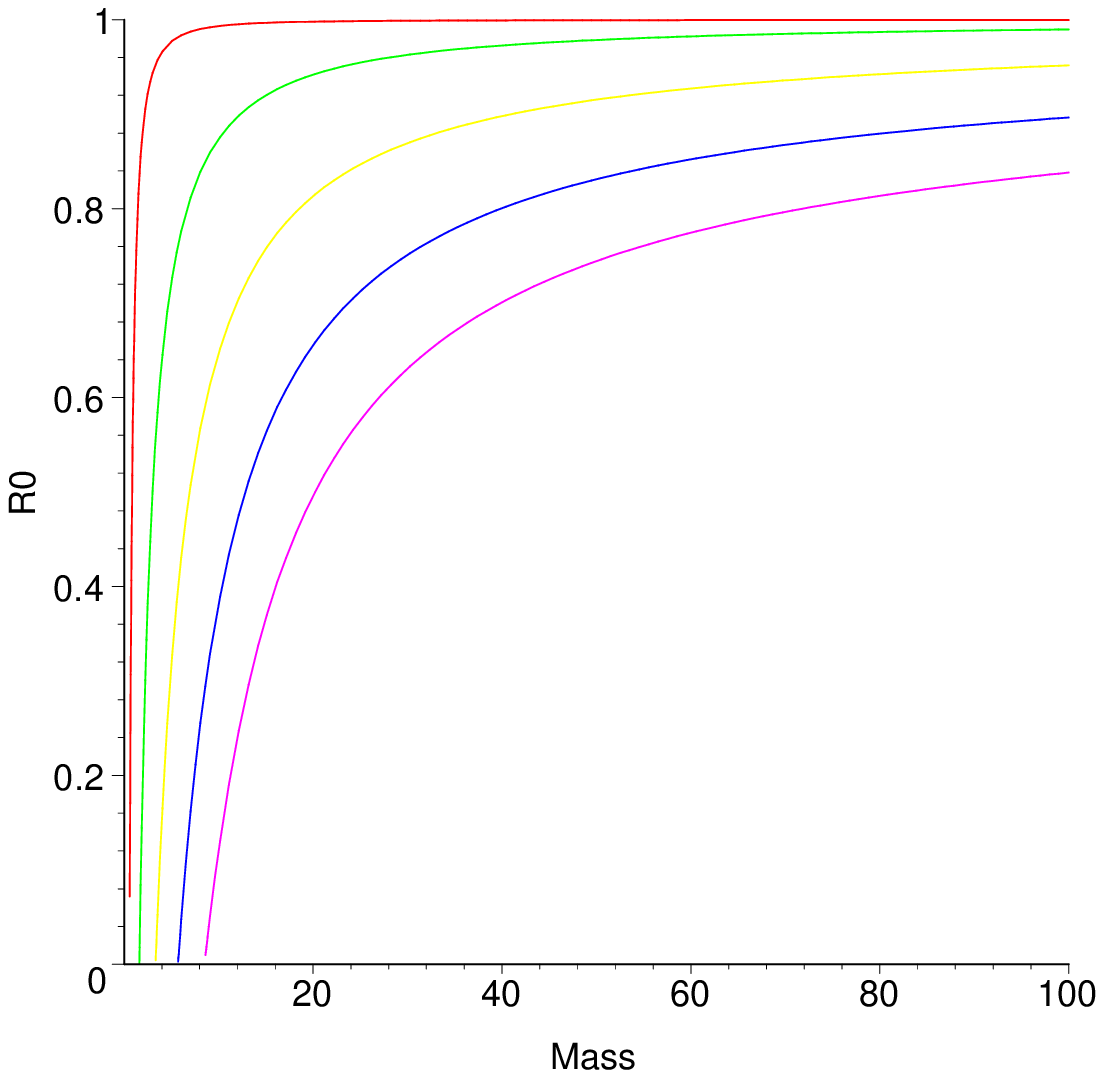}
\end{center}
{Figure 2: Ratio of the entropy with GUP to all orders in the Planck length
 to the entropy in the standard Hawking picture as a function of the BH hole mass (in units with $G=L=1$). From left to right,
$d=4,5,6,7,8$.}

\begin{center}
\includegraphics[height=8cm,
width=10cm]{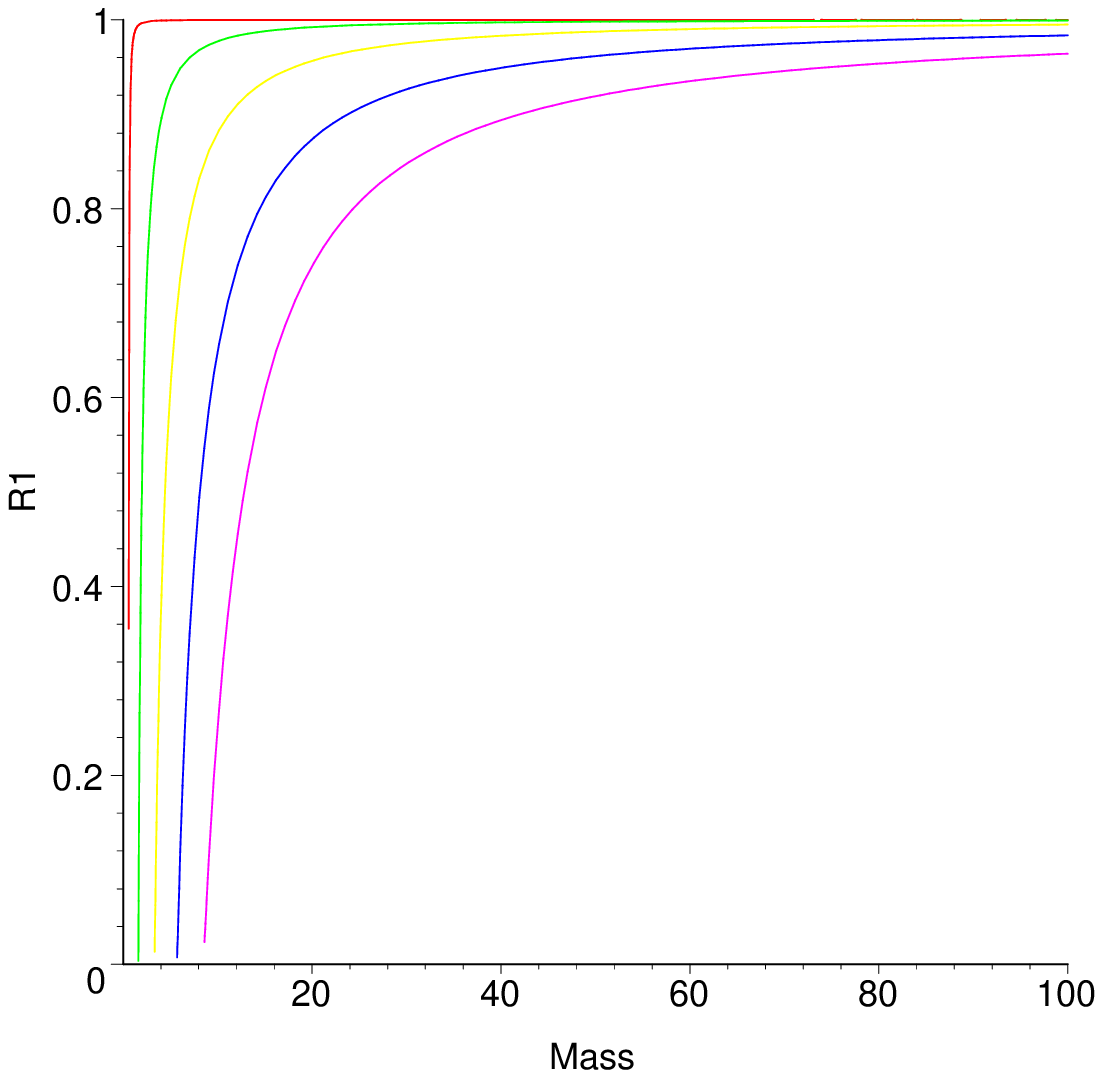}
\end{center}
{Figure 3: Ratio of the entropy with GUP to all orders in the Planck length
 to the entropy with GUP to leading order in the Planck length as a function of the BH hole mass (in units with $G=L=1$). From left to right,
$d=4,5,6,7,8$.}

\section{Black hole evaporation}

We consider now the mass loss rates and lifetimes of a BH of large
mass $m$. Once produced, the BH undergoes a number of phases before
completely evaporating or leaving an inert BH remnant in the
scenario with GUP. These phases are summarized in the following
\cite{bh5}

\textit{Balding phase}: During this phase, the BH lost hair
associated with multipole moments inherited from the initial
particles, and a fraction of the initial mass will be lost by
gravitational radiation.

\textit{Evaporation phase}: The BH starts losing its angular
momentum through the emission of Hawking radiation and possibly,
through super-radiance and undergoes emission of thermally
distributed quanta until the BH reaches the Planck scale. The
emitted spectrum contains all Standard model particles, which are
emitted on the brane, as well as gravitons, which are also emitted
in the bulk direction.

\textit{Planck phase}: During this phase, the semi classical picture
breaks down since the mass and/or the Hawking temperature approach
the Planck scale. Hence, a theory of quantum gravity is necessary to
study this phase. However, it is suggested that the BH will decay to
a few quanta with Planck-scale energies or to a inert remnant.

The usual thermodynamical description of the Hawking evaporation
process is usually performed with the canonical ensemble (CE)
approach. In the CE approach the energies of the emitted particles
are small compared to the BH mass. However, it was pointed out in
\cite{casadio}, that the CE approach is no longer appropriate in the
final stage of evaporation where BH is hot and its mass approaches
the Planck scale. Thus,  the correct description of the evaporation
process requires the use of the micro-canonical ensemble (MCE)
description.

In the following, ignoring the contribution of the grey-body factors,  we calculate the evaporation
rate of a massive BH such that $m/m_0$ is much greater than $O(1)$, where $m_0$ is the minimum BH mass allowed by the GUP.
In this approximation, the MCE corrections can be neglected and the energy density of the emitted particles in
(D+1)-dimensional space-time is given by
\begin{equation}
{\mathcal{E}}=2\Omega _{D-1}\int_{0}^{\infty }\frac{%
p^{D}e^{-\alpha ^{2}L_{Pl}^{2}p^2}}{e^{\beta p}-1}dp.
\end{equation}%
The evaluation of this integral proceeds by expanding the exponential and
the use of the following definition of the Riemann Zeta function
\begin{equation}
\int_{0}^{\infty }\frac{y^{s-1}}{e^{y}-1}dy=\Gamma (s)\zeta (s).
\end{equation}%
As a result we obtain%
\begin{equation}
{\mathcal{E}}=2\Omega _{D-1}T^{D+1}\sum_{n=0}^{\infty }\frac{%
(-1)^{n}}{n!}\left( \alpha L_{Pl}T_{H}\right) ^{2n}\Gamma \left(
2n+D+1\right) \zeta \left( 2n+D+1\right) .  \label{energy}
\end{equation}%
The series in Eq.(\ref{energy}) is an alternating series which converge when
$T_{H}<\alpha ^{-1}T_{Pl}$. However the existence of a maximum value of the
Hawking temperature implies a stronger condition on $T$. Using the expression of the Hawking maximum temperature given by Eq.$(\ref{Tmax})$, we have
\begin{equation}
\alpha \frac{T}{T_{Pl}}<\frac{d-3}{2\pi \sqrt{2}}.
\end{equation}
This constraint allows us to cut the series at $n=1$. Then we have
\begin{eqnarray}
{\mathcal{E}}&=&2\Omega _{D-1}T^{D+1}\Gamma \left( D+1\right) \zeta
\left( D+1\right)\times \nonumber \\
 & &\left( 1-\frac{(d-3)^{2}(D+1)(D+2)\zeta (D+3)}{8\pi
^{2}\zeta (D+1)}\left( \frac{T_{H}}{T_{H}^{max}}\right) ^{2}\right) .
\label{evap}
\end{eqnarray}%
Neglecting thermal emission in the bulk and assuming a $(D+1)$-dimensional brane, the intensity emitted by a massless scalar particle on
the brane is
\begin{equation}
\frac{dM}{dt}=-A_{D+1}{\mathcal{E}}\left( T_{H}\right) .
\label{em}
\end{equation}%
where $A_{D}=\Omega _{D-2}r_{c}^{D-2}$ is the horizon area of the induced BH
and $r_{c}=\left[ \frac{d-1}{2}\right] ^{1/(d-3)}\left[ \frac{d-1}{d-3}%
\right] ^{1/2}r_{h}$ is the critical radius of the BH considered as
an absorber \cite{bh3}. In Eq.(\ref{em}), the constancy of the
surface gravity over the horizon, allows to identify the Hawking
temperature of the higher dimensional BH as the temperature of the
induced BH on the brane.

Considering a four dimensional brane and using the corrected
Hawking temperature given by $\left(\ref{H-temp}\right) $ we obtain%
\begin{equation}
\frac{dm}{dt}=-\gamma_1 Ze^{-2W\left(-Z\right)}\left(1-\gamma_2 \alpha^2 Ze^{-W\left(-Z\right)}\right)
,  \label{rate}
\end{equation}
with $\gamma_1=\frac{\pi^2 e^2 M_{Pl}}{120\omega_d^{2}}(r_c/r_h)^2(d-3)^4m_0^{-2/(d-3)}$, $\gamma_2=
\frac{5e}{42\omega_d^2 }(d-3)^2m_0^{-2/(d-3)}$ and $Z=\frac{1}{e}\left(m_0/m\right)^{2/(d-3)}$. \\

In figure 4, we show the variation of the evaporation rate with the
BH mass. We observe that the evaporation phase ends when the BH mass
reaches the minimum mass $m_{0}$. In this case $Z=1/e$ and the
evaporation rate given by
\begin{equation}
\left( \frac{dm}{dt}\right) _{min}=-\gamma_1 e\left( 1-\gamma_2\alpha^2\right),
\end{equation}
is finite. It is important to note, that although we have used the
CE approach, the usual divergence at the end of the Hawking
evaporation in the standard description, is now completely removed
by the GUP. However, as pointed by several authors, the divergence
at the end of the Hawking evaporation process in the standard
description is a consequence of the incorrect use of the CE approach
and can be cured by the MCE treatment \cite{bh13,casadio}. In the
framework with GUP, the existence of a maximum temperature given by
$\left( \ref{Tmax}\right) $ suppress the evaporation process beyond
the Planck temperature. This behavior is similar to the prevention,
by the standard uncertainty principle, of the hydrogen
atom from total collapse. \\

Performing a Taylor expansion around the small parameter $Z$ we obtain
\begin{equation}
\frac{dm}{dt}=-\frac{\gamma_1}{e}(1-\gamma_2\alpha^2)\left(\frac{m_0}{m}\right)^{\frac{2}{d-3}}
-\frac{2\gamma_1}{e^2}(1-\frac{3}{2}\gamma_2\alpha^2)\left(\frac{m_0}{m}\right)^{\frac{4}{d-3}}+\cdots,
\end{equation}
\begin{center}
\includegraphics[height=8cm,
width=10cm]{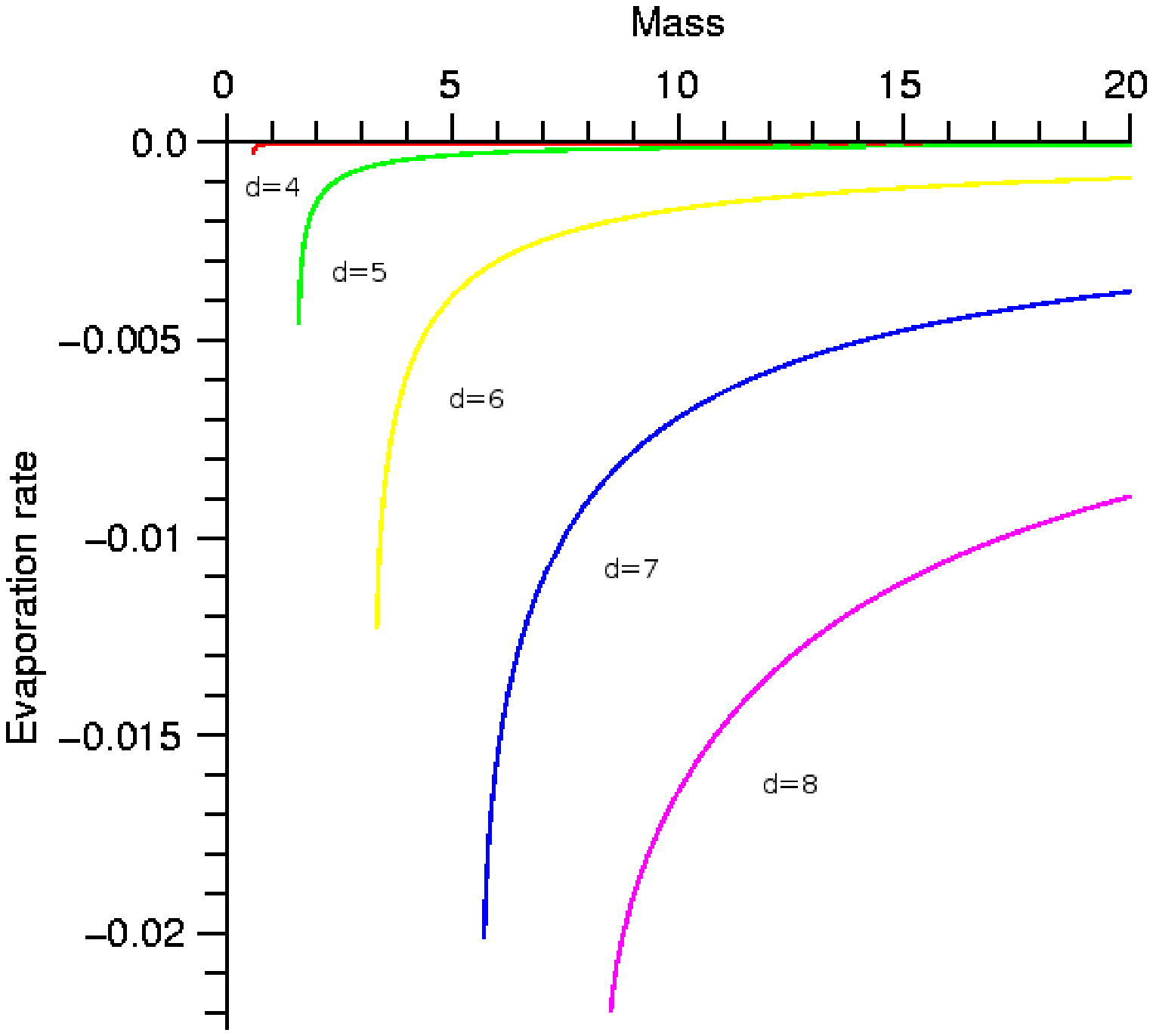}
\end{center}
\begin{center}
{\small Figure 4: The evaporation rate as a function of the black
hole mass (in units with $G=L=1$).}
\end{center}

In the framework with GUP, the Hawking evaporation process of BHs
with mass $m>m_{0}$
continue until the horizon radius becomes $\left( \delta X\right) _{min%
}$ leaving a Planck sized BH remnant. The nature of this BH remnant
$\left(
\hbox{BHR}\right) $ is best described by the specific heat. Using the definition $C_{d}=%
\frac{dM}{dT_{H}}$ we obtain
\begin{equation}
C_{d}=C_{d}^{0}\left(
1+W\left( -\frac{1}{e}\left( \frac{m_{0}}{m}\right) ^{^{2/(d-3)}}\right)
\right)\exp \left( \frac{1}{2}W\left( -\frac{1}{e}\left( \frac{m_{0}}{m}%
\right) ^{^{2/(d-3)}}\right) \right),
\end{equation}%
where $C_{d}^{0}=-4\pi \omega _{d}L_{{Pl}}m^{\frac{d-2}{d-3}}$ is the heat capacity without GUP. We observe that the heat capacity vanishes when $\left( 1+W\left( -\frac{1}{e%
}\left( \frac{m_{0}}{m}\right) ^{^{2/(d-3)}}\right) \right) =0$,
whose solution is given by $m=m_{0}$, corresponding to the end point
of the evaporation phase. This state, characterized by a maximal temperature, can be considered as
the ground state of the BH.  This interpretation is motivated by the fact
that the ground state is independent of the temperature \cite{ling}.
Thus, the vanishing of the specific heat and the entropy at the end of the
evaporation reveals, beside gravitational interaction with the
surrounding, the inert character of the BHRs and thus make them as
potential candidates for the origin of dark matter
\cite{Pisin,P-adler}.
We note that, as it is the case with the GUP to leading order in the Planck length widely used in the
literature, the BHRs seems to be also consequence of the GUP to all
orders in the Planck length considered in this paper
\cite{adler,bh25}. We note that BHRs can be found in different
contexts like noncommutative geometry \cite{spallucci,nico} and
effective models based on the \emph{Limiting Curvature Hypothesis}
(LCE) \cite{easson}. However in these scenarios the BH radiates
eternally. These models and others have in common that the
temperature of the BH  reaches a maximum before dropping  to zero. In
our framework, this behavior is forbidden  by the cut-off implemented by
the GUP. In figure 5, we show the variation of the specific heat
with the BH mass.

\begin{center}
\includegraphics[height=8cm,
width=10cm]{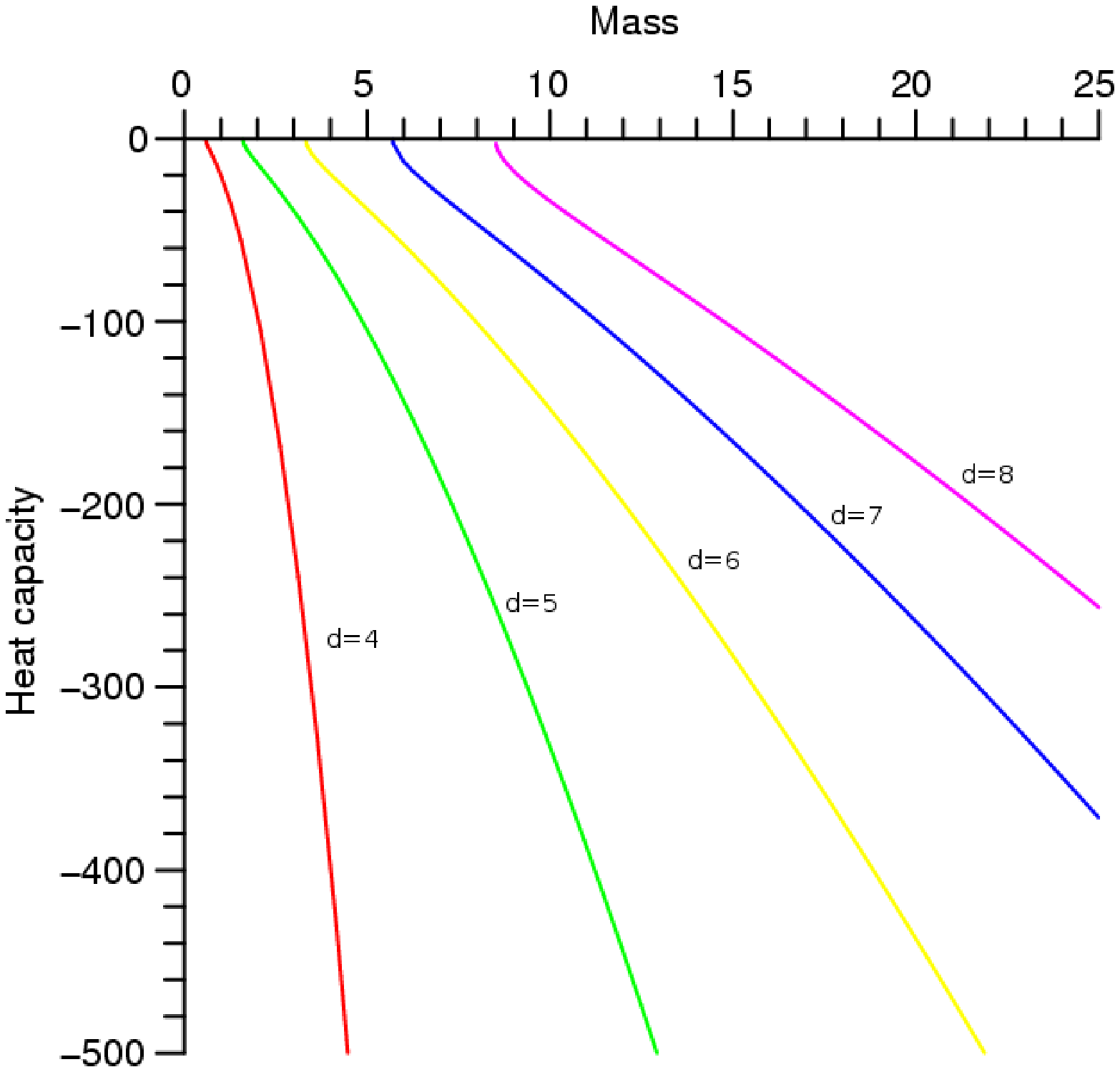}
\end{center}
\begin{center}
{\small Figure 5: Heat capacity as a function of the black hole mass
(in units with $G=L=1$).}
\end{center}

A Taylor expansion around $\frac{1}{e}(m_0/m)^{2/(d-3)}$ gives
\begin{equation}
C_{d}=C_{d}^{0}\left[
1-\frac{3}{2e}\left( \frac{m_{0}}{m}\right) ^{\frac{2}{d-3}}-\frac{7}{8e^{2}}%
\left( \frac{m_{0}}{m}\right) ^{\frac{4}{d-3}}-\frac{55}{6e^{3}}\left( \frac{%
m_{0}}{m}\right) ^{\frac{6}{d-3}}\right] .
\end{equation}%
For a BH with a mass larger than the minimum mass allowed by the
GUP, the heat capacity can be approximated by the standard
expression,  $C_{d}^{0}$. The corrections terms to the specific heat
due to GUP are all positive showing that the evaporation process is
accelerated, leading to a GUP corrected decay time smaller than the
decay time in the standard case.

Taking into account that the evaporation phase ends when the BH mass reaches $m_0$, we obtain from $\left( \ref{rate}\right) $, the following expression for the decay time
\begin{equation}
t_{d}=-\frac{(d-3)m_0}{2\gamma_1 e^{(d-3)/2}}\left(I\left(d+1,d-3,Z_i\right)+\gamma_2\alpha^2I\left(d-1,d-3,Z_i\right)\right),\label{time}
\end{equation}
where $Z_i=-\frac{1}{e}(m_0/m_i)^{\frac{2}{d-3}}$, $m_i$ the initial
mass of the BH and
\begin{equation}
I\left(p,q,Z_i\right))=-\left(-1\right)^{\frac{q}{2}}\int_{Z_i}^{-\frac{1}{e}}
W(Z)^{-\frac{p}{2}}e^{-\frac{q}{2}W(Z)}dZ,\label{int}
\end{equation}
These integrals can be evaluated analytically in terms of the Lambert Function $W(x)$ and the Whittaker function $M(\alpha;\beta;x)$.

A plot of the decay time as a function of the BH mass obtained from
a exact evaluation of the integrals in Eq.$(\ref{int})$ in shown in
Fig. 5. We observe that the decay time is a rapidly decreasing
function of the space-time dimension. This confirm the fact that BHs
in higher dimensional space-times are hotter and decay faster.
\begin{center}
\includegraphics[height=8cm,
width=10cm]{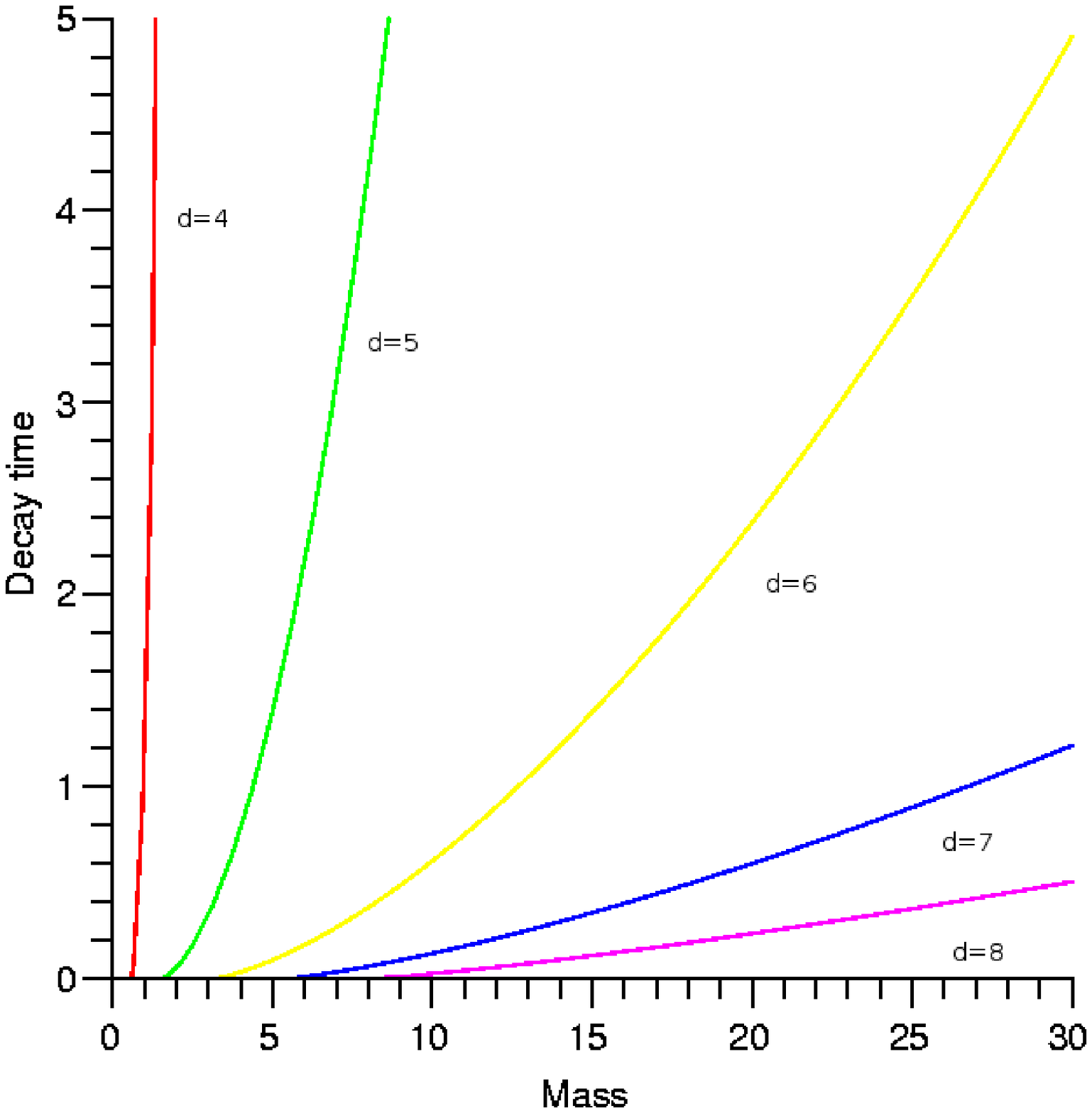}
\end{center}
\begin{center}
{Figure 6: Decay time as a function of the black hole mass (in units
with $G=L=1$).}
\end{center}

In the rest of this section, we proceed to a comparison of our
results with the ones obtained in the framework of the GUP to
leading order in the Planck length \cite{noza,bh25}. From the
saturate GUP defined by Eq.$(\ref{gup})$ we obtain
\begin{equation}
(\delta P)=\frac{\delta X}{\alpha^2 L_{Pl}^2}\left(1-\sqrt{1-\frac{\alpha^2 L_{Pl}^2}{(\delta X)^2}}\right).
\end{equation}
This leads to the minimal length given by Eq.(\ref{l0}).
Following the same calculations as above, we obtain
\begin{equation}
\label{T1}
T_{H}=\frac{d-3}{2\pi\alpha}T_{Pl}Z^{-1}\left(1-\sqrt{1-Z^{2}}\right),
\end{equation}
\begin{equation}
C_d=2\pi\alpha m_0Z^{4-d}\frac{\sqrt{1-Z^2}}{\sqrt{1-Z^2}-1}.
\end{equation}
with $Z=(m_0/m)^{1/(d-3)}$ and $m_0=(\alpha/\omega_d)^{d-3}$ is the minimum BH mass allowed by the GUP. Substituting $Z=1$ in Eq.(\ref{T1}), we obtain the maximum BH temperature $T_{H}^{max}=\frac{d-3}{2\pi\alpha}T_{Pl}$.

The calculation of the entropy gives
\begin{equation}
S_d=-\frac{(d-2)(d-3)}{16 L_{Pl}^{d-2}}A_0\int_{1}^{(A_0/A)^{2/(d-2)}}\frac{y^{\frac{2-d}{2}}}{1-\sqrt{1-y}}dy,
\end{equation}
where $y=\left(A_0/A\right)^{2/(d-2}$. The integral can be evaluated for given values of $d$ and we obtain for $d=4,5,6$ the following expressions
\begin{equation}
S_4=-\frac{A_0}{8L_{Pl}^2}\left[\frac{1}{2}\hbox{ln}\frac{\sqrt{1-y}+1}{\sqrt{1-y}-1}+\frac{1}{\sqrt{1-y}-1}\right]_{y=A_0/A},
\end{equation}
\begin{equation}
S_5=-\frac{A_0}{8L_{Pl}^3}\sqrt{\frac{1-y}{y}}\left[1-\frac{1}{y}-\frac{1}{y\sqrt{1-y}}\right]_{y=(A_0/A)^{2/3}},
\end{equation}
\begin{equation}
S_6=-\frac{A_0}{16L_{Pl}^2}\left[\frac{1}{2}\hbox{ln}\frac{\sqrt{1-y}+1}{\sqrt{1-y}-1}-\frac{1}{\sqrt{1-y}+1}
-\frac{1}{(\sqrt{1-y}-1)^2}\right]_{y=(A_0/A)^{1/2}}.
\end{equation}
A Taylor expansion in the parameter $y$ gives again the log area correction in the case of even dimension.

The calculation of the evaporation rate in the framework with GUP to
leading order in the Planck length requires a careful analysis. The
calculations done until now used the usual momentum measure in the
derivation of the Stefan-Boltzmann law. However, it is easy to show,
following \cite{ke}, that the fundamental cell in the momentum space
is squeezed by the presence of the minimal and becomes
$dp/(1+\alpha^2 L_{Pl}^2 p^2)$, like the momentum measure given by
Eq.$(\ref{measure})$ in our framework. Then, using the Bose-Einstein
statistic, the energy density of the emitted particles is given in
the CE approach by
\begin{equation}
{\mathcal{E}}=2\Omega _{D-1}\int_{0}^{\infty }
\frac{p^{D}}{(1+\alpha^2 L_{Pl}^2 p^2)e^{\beta p}-1}dp.
\end{equation}
We then arrive to the same leading contribution given by Eq.(\ref{evap}), with the  condition on the Hawking temperature given now by
\begin{equation}
\frac{T_{H}}{T_{Pl}}<\frac{d-3}{2\pi}.
\end{equation}
Using the definition of the evaporation rate given by Eq.(\ref{em})
we obtain
\begin{equation}
\frac{dm}{dt}=-\alpha_1Z^{-3}\left(1-\sqrt{1-Z}\right)^4\left[1-\alpha_2 Z^{-1}\left(1-\sqrt{1-Z}\right)^{2}\right],
\label{e-r-fo}
\end{equation}
where $Z=\left(m_0/m\right)^{2/(d-3)}, \alpha_1=\frac{2^{(d-5)/(d-3)}\pi^2}{15\alpha^4}{M_{Pl}\omega_d^2m_0^{2/(d-3)}(d-3)^3}{(d-1)^{\frac{d-1}{d-3}}},  \alpha_2=\frac{10}{21}(d-3)^2$.

The Hawking evaporation ends when the BH mass becomes equal to $m_0$
with a minimum evaporation rate given by
\begin{equation}
\left(\frac{dm}{dt}\right)_{min}=-\alpha_1\left(1-\alpha_2\right).
\end{equation}
The expression of the decay time follows from Eq.(\ref{e-r-fo}) and is given by
\begin{equation}
t=\frac{d-3}{2\alpha_1}m_0\left( I\left(-4,\frac{7-d}{2},Z_i\right)
+\alpha_2 I\left(-2,\frac{5-d}{2},Z_i\right)\right),
\label{t1}
\end{equation}
where
\begin{equation}
 I\left(p,q,Z_i\right)=\int_{Z_i}^{1} dZ Z^{q}\left(1-\sqrt{1-Z}\right)^p.
\end{equation}
The evaluation of the integrals gives complicated and long expressions to be presented here. Our results  differ from the ones obtained in \cite{bh25,noza} by the presence of the seconds terms in Eqs.(\ref{e-r-fo},\ref{t1}), which are a consequence of the squeezing of the momentum space.

Before ending this section we consider the  multiplicity of
particles emitted during the evaporation process. Assuming that the
BH radiates mainly on the 3-brane, and ignoring the grey body factors
the multiplicity is given by \cite{marco}
\begin{equation}
N=\frac{d-3}{d-2}\frac{\zeta(3)}{3\zeta(4)}\frac{M}{T_H},
\end{equation}%
In the case of the GUP to all orders in the Planck length and with
the BH temperature given by $\left( \ref{H-temp}\right) $, we
observe that, compared to the standard picture, the additional
exponential factor leads to a reduction of the average multiplicity
with increasing number of extra dimensions.

In table 1, we show the corrected thermodynamics parameters of two
five dimensional BHs with initial mass equal to $5M_{Pl}$ and
$10M_{Pl}$ in the frameworks on the GUP to leading order
(GUP$^{lo}$) and the GUP to all orders in the Planck length
(GUP$^{ao}$). The first row represents the semiclassical results
obtained with the Heisenberg uncertainty principle (HUP). We observe
a reduction of the entropy and the decay time in our framework. This
reduction becomes significant as the number of extra dimensions
increases. As a result, the multiplicity of emitted particles in BH
decay is then significantly suppressed. For example, for $d=8$ and
$M=9 M_{Pl}$ the multiplicity with GUP$^{ao}$ is suppressed by a
factor $-12\%$ in comparison
 to the multiplicity with GUP$^{fo}$ and by $-33\%$ in comparison to the standard multiplicity. For $d=8$ and $M=12 M_{Pl}$, the suppression factors are respectively $-5\%$ and $-24\%$. The
reduction of the entropy indicates the breakdown of the
semiclassical picture and that BH in the framework with a
GUP have less degrees of freedom compared to the standard picture.
As it is the case with the GUP$^{lo}$, the effects of the GUP
becomes stronger as the minimum BH mass increases.  These results
may have important consequences on possible BHRs production in
particle colliders and in ultrahigh energy cosmic ray (UHECR)
airshowers.
\bigskip

\noindent Table 1. {Corrected thermodynamic parameters for two
five-dimensional BHs with mass 5 and 10 (in Planck units). The
deviations from the results with GUP$^{lo}$ are also given.}
\begin{center}
M=5

\begin{tabular}{|c|c|c|c|c|c|c|}
\hline
$ $ & $m_0$ & $T_{i}$ &
$T_{f}$ & $S$ & $t$ & $N$\\ \hline
${HUP}$ & ${ -}$ &  0.077 &
${ \infty }$ &
 86.29 &  2.01 & 12 \\ \hline
${ GUP^{lo}}$ & 1.178 &  0.082
&  0.318 &  33.52 (-61$\%$)  &  1.47 (-27$\%$)%
&11.22 (-6.3$\%$) \\ \hline
${GUP^{ao}}$ &  1.60  &  0.082
 &  0.22 &  %
15.56 (-82$\%$)  &  0.51 ( -75$\%$) &11.20 (-6.5$\%$) \\
\hline
\end{tabular}

\bigskip

M=10

\begin{tabular}{|c|c|c|c|c|c|c|}
\hline ${ }$ & $m_0$ & $T_i$ & {$T_{f}$} & $S$ & $t$ & $N$ \\ \hline
${HUP}$ & ${-}$ & 0.054 & ${\infty }$ & 244.07 & 8.06 & 33.88 \\
\hline ${GUP^{lo}}$ & 1.178 & 0.056  & 0.318  &109.11 (-55$\%$) &
7.007 (-13$\%$) & 32.85 (-3$\%$)  \\ \hline
${GUP^{ao}}$ & 1.60 & 0.056 & 0.22 & 53.32 (-78$\%$) & 2.53 (-69$\%$)& 32.83 (-3$\%$)   \\
\hline
\end{tabular}
\end{center}

\section{Conclusion}

We have considered, in the scenario with large extra dimensions and
with a GUP to all orders in the Planck length, the corrections to
the BHs thermodynamic parameters. We have obtained exact expressions
for the Hawking temperature and entropy. We have also reproduced the
log-area logarithmic corrections of the entropy in the case of even
number of extra dimensions.  Using the canonical ensemble
description, we have investigated the Hawking evaporation process of a
large BH and shown that the end of the process is a black hole
remnant (BHR) with zero entropy, zero heat capacity and a finite non
zero temperature. We have shown that BHs in the framework of a GUP
to all orders in the Planck length have less degrees of freedom, are
hotter and decay faster than in the Hawking semiclassical picture
and in the framework of the GUP to leading order in the Planck
length.
\bigskip

\section*{Acknowledgments}{The author thanks the Algerian Ministry of Scientific Research and High Education for financial support and Professor W. Greiner at FIAS Frankfurt for warm hospitality.}

\end{document}